\documentclass[10pt]{article}
\usepackage{latexsym,graphicx}
\newcommand{\be}{\begin{equation}}
\newcommand{\ee}{\end{equation}}
\usepackage[left=2.5cm,top=2.5cm,right=2.5cm,bottom=1.5cm]{geometry}
\def\n{\noindent}
\catcode `\@=11
\catcode `\@=12
\begin{document}
\begin{center}
\large{\bf {String Cosmological Model in Cylindrically Symmetric Inhomogeneous Universe with Electromagnetic Field}} \\
\vspace{10mm}
\normalsize{Anirudh Pradhan $^1$, Anju Rai $^2$ and Raj Bali $^3$ }\\
\vspace{5mm}
\normalsize{$^{1}$ Department of Mathematics, G. L. A. University, Mathura-281 406, Uttar Pradesh, India \\
\vspace{5mm}
E-mail: pradhan@iucaa.ernet.in} \\
\vspace{5mm}
\normalsize{$^{2}$ Department of Mathematics, Keystone Group of Institutions, Pilod, Jhunjhinu-333 029, Rajasthan, India \\
\vspace{5mm}
E-mail: anjuraianju@yahoo.co.in} \\
\vspace{5mm}
\normalsize{$^{3}$Department of Mathematics, University of Rajasthan, Jaipur-302 004, Rajasthan, India \\
\vspace{5mm}
E-mail: balir5@yahoo.co.in}\\ 
\vspace{5mm}
\end{center}
\vspace{10mm}
\begin{abstract} 
Cylindrically symmetric inhomogeneous string cosmological models in presence of 
electromagnetic field is investigated. We have assumed that $F_{12}$ is the only 
non-vanishing component of $F_{ij}$. The Maxwell's equations require that $F_{12}$  
is the function of $x$ and $t$ both and the magnetic permeability is the function 
of $x$ and $t$ both. To get the deterministic solution, it has been assumed that 
the expansion ($\theta$) in the model is proportional to the eigen value 
$\sigma^{1}~~_{1}$ of the shear tensor $\sigma^{i}~~_{j}$. The derived model represents 
the inflationary scenario as the proper volume increases exponentially with cosmic time. 
It is observed that the model has a point type singularity. The physical and geometric 
aspects of the model are also discussed.   
\end{abstract}
\smallskip
\n Keywords : Cosmic string, Electromagnetic field, Inhomogeneous universe\\
\n PACS number: 98.80.Cq, 04.20.-q 
\section{Introduction}
Cosmic strings play an important role in the study of the early universe. These strings 
arise during the phase transition after the big bang explosion as the temperature goes 
down below some critical temperature as predicted by grand unified theories 
\cite{ref1}${-}$ \cite{ref5}. It is believed that cosmic strings give rise to density 
perturbations which lead to formation of galaxies \cite{ref6}. These cosmic strings have 
stress energy and couple to the gravitational field. Therefore, it is interesting to study 
the gravitational effect which arises from strings. The general treatment of strings was 
initiated by Letelier \cite{ref7,ref8} and Stachel \cite{ref9}. The occurrence of magnetic 
fields on galactic scale is well-established fact today, and their importance for a variety 
of astrophysical phenomena is generally acknowledged as pointed out Zel'dovich \cite{ref10}. 
Also Harrison \cite{ref11} has suggested that magnetic field could have a cosmological origin. 
As a natural consequences, we should include magnetic fields in the energy-momentum tensor of the 
early universe. The choice of anisotropic cosmological models in Einstein system of field 
equations leads to the cosmological models more general than Robertson-Walker model \cite{ref12}. 
The presence of primordial magnetic fields in the early stages of the evolution of the universe 
has been discussed by several authors (Misner, Thorne and Wheeler \cite{ref13}; Asseo and Sol 
\cite{ref14}; Pudritz and Silk \cite{ref15}; Kim, Tribble, and Kronberg \cite{ref16}; Perley and 
Taylor \cite{ref17}; Kronberg, Perry and Zukowski \cite{ref18}; Wolfe, Lanzetta and Oren \cite{ref19}; 
Kulsrud, Cen, Ostriker and Ryu \cite{ref20}; Barrow \cite{ref21}). Melvin \cite{ref22}, in his 
cosmological solution for dust and electromagnetic field suggested that during the evolution 
of the universe, the matter was in a highly ionized state and was smoothly coupled with the field, 
subsequently forming neutral matter as a result of universe expansion. Hence the 
presence of magnetic field in string dust universe is not unrealistic. \\

Benerjee et al. \cite{ref23} have investigated an axially symmetric Bianchi type I string dust 
cosmological model in presence and absence of magnetic field using a supplementary 
condition $\alpha = a \beta$ between metric potential where $\alpha = \alpha(t)$ and 
$\beta = \beta(t)$ and $a$ is constant. The string cosmological models with a magnetic 
field are also discussed by Chakraborty \cite{ref24}, Tikekar and Patel \cite{ref25,ref26}. Patel 
and Maharaj \cite{ref27} investigated stationary rotating world model with magnetic field. Ram 
and Singh \cite{ref28} obtained some new exact solution of string cosmology with and without a 
source free magnetic field for Bianchi type I space-time in the different basic form 
considered by Carminati and McIntosh \cite{ref29}. Singh and Singh \cite{ref30} investigated string 
cosmological models with magnetic field in the context of space-time with $G_{3}$ 
symmetry. Singh \cite{ref31} has studied string cosmology with electromagnetic fields in 
Bianchi type-II, -VIII and -IX space-times. Lidsey, Wands and Copeland \cite{ref32} have 
reviewed aspects of super string cosmology with the emphasis on the cosmological 
implications of duality symmetries in the theory. Bali et al. \cite{ref33,ref34,ref35} have 
investigated Bianchi type I magnetized string cosmological models.\\

Cylindrically symmetric space-time play an important role in the study of the universe 
on a scale in which anisotropy and inhomogeneity are not ignored. Inhomogeneous 
cylindrically symmetric cosmological models have significant contribution in 
understanding some essential features of the universe such as the formation of 
galaxies during the early stages of their evolution. Bali and Tyagi \cite{ref36} and 
Pradhan et al. \cite{ref37,ref38} have investigated cylindrically symmetric inhomogeneous 
cosmological models in presence of electromagnetic field. Barrow and Kunze \cite{ref39,ref40} 
found a wide class of exact cylindrically symmetric flat and open inhomogeneous string 
universes. In their solutions all physical quantities depend on at most one space 
coordinate and the time. The case of cylindrical symmetry is natural because of the 
mathematical simplicity of the field equations whenever there exists a direction in 
which the pressure equal to energy density. \\

Recently Baysal et al. \cite{ref41} have investigated some string cosmological models in 
cylindrically symmetric inhomogeneous universe. Motivated by the situation discussed 
above, in this paper, we have generalized these solutions by including electromagnetic 
field tensor. We have taken string and electromagnetic field together as the source 
gravitational field as magnetic field are anisotropic stress source and low strings 
are one of anisotropic stress source as well. This paper is organized as follows: 
The metric and field equations are presented in Section $2$. In Section $3$, we deal 
with the solution of the field equations in presence of electromagnetic field with perfect 
fluid distribution. In Section $4$, we have given the concluding remarks.  
\section{The Metric and Field Equations}
We consider the metric in the form 
\begin{equation}
\label{eq1}
ds^{2} = A^{2}(dx^{2} - dt^{2}) + B^{2} dy^{2} + C^{2} dz^{2},
\end{equation}
where $A$, $B$ and $C$ are functions of $x$ and $t$.
The energy momentum tensor for the string with electromagnetic field 
has the form 
\begin{equation}
\label{eq2}
T^{j}_{i} = \rho u_{i}u^{j} - \lambda x_{i}x^{j} +  E^{j}_{i},
\end{equation}
where $u_{i}$ and $x_{i}$ satisfy conditions
\begin{equation}
\label{eq3}
u^{i} u_{i} = - x^{i} x_{i} = -1, ~ \mbox{and} ~ u^{i} x_{i} = 0.
\end{equation}
Here $\rho$ being the rest energy density of the system of strings, $\lambda$ the 
tension density of the strings, $x^{i}$ is a unit space-like vector representing 
the direction of strings so that $x^{2} = 0 = x^{3} = x^{4}$ and $x^{1} \ne 0$, 
and $u^{i}$ is the four velocity vector satisfying the 
following conditions
\begin{equation}
\label{eq4}
g_{ij} u^{i} u^{j} = -1.
\end{equation}
In Eq. (\ref{eq2}), $E^{j}_{i}$ is the electromagnetic field given by Lichnerowicz \cite{ref42} 
\begin{equation}
\label{eq5}
E^{j}_{i} = \bar{\mu}\left[h_{l}h^{l}\left(u_{i}u^{j} + \frac{1}{2}g^{j}_{i}\right) 
- h_{i}h^{j}\right],
\end{equation}
where $\bar{\mu}$ is the magnetic permeability and $h_{i}$ the magnetic flux vector 
defined by
\begin{equation}
\label{eq6}
h_{i} = \frac{1}{\bar{\mu}} \, {^*}F_{ji} u^{j},
\end{equation}
where the dual electromagnetic field tensor $^{*}F_{ij}$ is defined by Synge \cite{ref43} 
\begin{equation}
\label{eq7}
^{*}F_{ij} = \frac{\sqrt{-g}}{2} \epsilon_{ijkl} F^{kl}.
\end{equation}
Here $F_{ij}$ is the electromagnetic field tensor and $\epsilon_{ijkl}$ is the Levi-Civita 
tensor density. \\
The components of electromagnetic field are obtained as
\[
 E^{1}_{1} = E^{2}_{2} = E^{4}_{4} =\frac{F_{12}^{2}}{2\bar{\mu} A^{2}B^{2}} ,
\]
\begin{equation}
\label{eq8}
  E^{3}_{3} =  - \frac{F_{12}^{2}}{2\bar{\mu} A^{2}B^{2}}.
\end{equation}
In the present scenario, the comoving coordinates are taken as 
\begin{equation}
\label{eq9}
u^{i} = \left(0, 0, 0, \frac{1}{A}\right). 
\end{equation}
We choose the direction of string parallel to x-axis so that
\begin{equation}
\label{eq10}
x^{i} = \left(\frac{1}{A}, 0, 0, 0 \right). 
\end{equation}
We consider that $F_{12}$ is the only non-vanishing component of $F_{ij}$ so that $h_{3} \ne 0$. Maxwell's equations  
\begin{equation}
\label{eq11}
F_[ij;k] = 0,
\end{equation}
\begin{equation}
\label{eq12}
\left[\frac{1}{\bar{\mu}}F^{ij}\right]_{;j} = 0,
\end{equation}
require that $F_{12}$ is the function of $x$ and $t$ both and the magnetic permeability is also the 
functions of $x$ and $t$ both. The semicolon represents a covariant differentiation. 

The Einstein's field equations (with $\frac{8\pi G}{c^{4}} = 1$) 
\begin{equation}
\label{eq13}
R^{j}_{i} - \frac{1}{2} R g^{j}_{i}  = - T^{j}_{i},
\end{equation}
for the line-element (\ref{eq1}) lead to the following system of equations:  
\[
\frac{B_{44}}{B} + \frac{C_{44}}{C} - \frac{A_{4}}{A}\left(\frac{B_{4}}{B} + 
\frac{C_{4}}{C}\right) - \frac{A_{1}}{A}\left(\frac{B_{1}}{B} + \frac{C_{1}}{C}\right) 
-\frac{B_{1}C_{1}}{BC}  + \frac{B_{4} C_{4}}{B C} 
\]
\begin{equation}
\label{eq14}
= \left[ \lambda - \frac{F^{2}_{12}}{2\bar{\mu} A^{2} B^{2}} \right] A^{2},
\end{equation}
\begin{equation}
\label{eq15}
\left(\frac{A_{4}}{A}\right)_{4} - \left(\frac{A_{1}}{A}\right)_{1} + \frac{C_{44}}{C} - 
\frac{C_{11}}{ C} =  - \left[\frac{F^{2}_{12}}{2\bar{\mu} A^{2} B^{2}}\right] A^{2},
\end{equation}
\begin{equation}
\label{eq16}
\left(\frac{A_{4}}{A}\right)_{4} - \left(\frac{A_{1}}{A}\right)_{1} + \frac{B_{44}}{B} - 
\frac{B_{11}}{B} =   \left[\frac{F^{2}_{12}}{2\bar{\mu} A^{2} B^{2}}\right] A^{2},
\end{equation}
\[
- \frac{B_{11}}{B} - \frac{C_{11}}{C} + \frac{A_{1}}{A}\left(\frac{B_{1}}{B} + \frac{C_{1}}
{C}\right) + \frac{A_{4}}{A}\left(\frac{B_{4}}{B} + \frac{C_{4}}{C}\right) -
\frac{B_{1}C_{1}}{BC}  + \frac{B_{4} C_{4}}{B C} 
\]
\begin{equation}
\label{eq17}
= \left[\rho - \frac{F^{2}_{12}}{2\bar{\mu} A^{2} B^{2}}\right] A^{2},
\end{equation}
\begin{equation}
\label{eq18}
\frac{B_{14}}{B} + \frac{C_{14}}{C} - \frac{A_{4}}{A}\left(\frac{B_{1}}{B} + \frac{C_{1}}{C}
\right) - \frac{A_{1}}{A}\left(\frac{B_{4}}{B} + \frac{C_{4}}{C}\right) = 0,
\end{equation}
where the sub indices $1$ and $4$ in A, B, C and elsewhere denote ordinary differentiation
with respect to $x$ and $t$ respectively.

The velocity field $u^{i}$ is irrotational. The scalar expansion $\theta$, shear scalar 
$\sigma^{2}$, acceleration vector $\dot{u}_{i}$ and proper volume $V^{3}$ are respectively 
found to have the following expressions:
\begin{equation}
\label{eq19}
\theta = u^{i}_{;i} = \frac{1}{A}\left(\frac{A_{4}}{A} + \frac{B_{4}}{B} + \frac{C_{4}}{C}
\right),
\end{equation}
\begin{equation}
\label{eq20}
\sigma^{2} = \frac{1}{2} \sigma_{ij} \sigma^{ij} = \frac{1}{3}\theta^{2} - \frac{1}{A^{2}}
\left(\frac{A_{4}B_{4}}{AB} + \frac{B_{4}C_{4}}{BC} + \frac{C_{4}A_{4}}{CA}\right),
\end{equation}
\begin{equation}
\label{eq21}
\dot{u}_{i} = u_{i;j}u^{j} = \left(\frac{A_{1}}{A}, 0, 0, 0\right) 
\end{equation}
\begin{equation}
\label{eq22}
V^{3} = \sqrt{-g} = A^{2} B C,
\end{equation}
where $g$ is the determinant of the metric (\ref{eq1}). Using the field equations and 
the relations (\ref{eq19}) and (\ref{eq20}) one obtains the Raychaudhuri's equation as
\begin{equation}
\label{eq23}
\dot{\theta} = \dot{u}^{i}_{;i} - \frac{1}{3}\theta^{2} - 2 \sigma^{2} - \frac{1}{2} 
\rho_{p},
\end{equation}
where dot denotes differentiation with respect to $t$ and
\begin{equation}
\label{eq24}
R_{ij}u^{i}u^{j} = \frac{1}{2}\rho_{p}.
\end{equation}
 With the help of Eqs. (\ref{eq1})$-$ (\ref{eq3}), (\ref{eq9}) and (\ref{eq10}), the 
Bianchi identity $\left(T^{ij}_{;j}\right)$ reduced to two equations:
\begin{equation}
\label{eq25}
\rho_{4} - \frac{A_{4}}{A}\lambda + \left(\frac{A_{4}}{A} + \frac{B_{4}}{B} + 
\frac{C_{4}}{C}\right)\rho = 0
\end{equation}
and
\begin{equation}
\label{eq26}
\lambda_{1} - \frac{A_{1}}{A}\rho + \left(\frac{A_{1}}{A} + \frac{B_{1}}{B} + 
\frac{C_{1}}{C}\right)\lambda = 0.
\end{equation}
Thus, due to all the three (strong, weak and dominant) energy conditions, one finds 
$\rho \geq 0$ and $\rho_{p} \geq 0$, together with the fact that the sign of $\lambda$ 
is unrestricted, it may take values positive, negative or zero as well.  
\section{Solutions of the Field Equations}
As in the case of general-relativistic cosmologies, the introduction of inhomogeneities 
into the string cosmological equations produces a considerable increase in mathematical 
difficulty: non-linear partial differential equations must now be solved. In practice, 
this means that we must proceed either by means of approximations which render the non-
linearities tractable, or we must introduce particular symmetries into the metric of the 
space-time in order to reduce the number of degrees of freedom which the inhomogeneities 
can exploit. \\
To get a determinate solution, let us assume that expansion ($\theta$) in the model 
is proportional to the value $\sigma^{1}~~_{1}$ of the shear tensor 
$\sigma^{i}~~_{j}$. This condition leads to
\begin{equation}
\label{eq27}
A = (BC)^{n},
\end{equation}
where $n$ is a constant. Equations (\ref{eq15}) and (\ref{eq16}) lead to
\begin{equation}
\label{eq28}
\frac{F^{2}_{12}}{\bar{\mu} B^{2}} = \frac{B_{44}}{B} - \frac{B_{11}}{B} - \frac{C_{44}}{C} 
+ \frac{C_{11}}{C}.
\end{equation}
and
\begin{equation}
\label{eq29}
2\left(\frac{A_{4}}{A}\right)_{4} - 2\left(\frac{A_{1}}{A}\right)_{1} + \frac{B_{44}}{B} - 
\frac{B_{11}}{B} + \frac{C_{44}}{C} - \frac{C_{11}}{ C} = 0.
\end{equation}
Using (\ref{eq27}) in (\ref{eq18}) reduces to
\begin{equation}
\label{eq30}
\frac{B_{41}}{B} + \frac{C_{41}}{C} - 2n \left(\frac{B_{4}}{B} + \frac{C_{4}}{C}\right)
\left(\frac{B_{1}}{B} + \frac{C_{1}}{C}\right) = 0.
\end{equation}
To get the deterministic solution, we assume 
\begin{equation}
\label{eq31}
B = f(x)g(t) ~ ~ \mbox{and} ~ ~ C = h(x) k(t)
\end{equation}
and discuss its consequences below in this paper.

In this case Eq. (\ref{eq30}) reduces to 
\begin{equation}
\label{eq32}
\frac{f_{1}/f}{h_{1}/h} = - \frac{(2n - 1)(k_{4}/k) + 2n(g_{4}/g)}{(2n - 1)(g_{4}/g) + 
2n(k_{4}/k)} = K \mbox{(constant)}.
\end{equation}
which leads to
\begin{equation}
\label{eq33}
\frac{f_{1}}{f} = K\frac{h_{1}}{h},
\end{equation}
and
\begin{equation}
\label{eq34}
\frac{k_{4}/k}{g_{4}/g} = \frac{K - 2nK - 2n}{2nK + 2n - 1} = a \mbox{(constant)}.
\end{equation}
From Eqs. (\ref{eq33}) and (\ref{eq34}), we obtain
\begin{equation}
\label{eq35}
f = \alpha h^{K},
\end{equation}
and
\begin{equation}
\label{eq36}
k = \delta g^{a},
\end{equation}
where $\alpha$ and $\delta$ are integrating constants.

From Eqs. (\ref{eq29}) and (\ref{eq27}), we obtain
\[
(2n + 1)\frac{B_{44}}{B} - 2n \frac{B^{2}_{4}}{B^{2}} + (2n + 1)\frac{C_{44}}{C} - 
2n\frac{C^{2}_{4}}{C^{2}} =
\]
\begin{equation}
\label{eq37}
(2n + 1)\frac{B_{11}}{B} + (2n + 1)\frac{C_{11}}{C} - 2n \frac{B^{2}_{1}}{B^{2}} - 2n \frac{C^{2}_{1}}
{C^{2}} = \mbox{N (constant)}.
\end{equation}
Eqs. (\ref{eq31}) and (\ref{eq37}) lead to
\begin{equation}
\label{eq38}
gg_{44} + r g^{2}_{4} = s g^{2},
\end{equation}
where 
$$
r = \frac{a(a - 1) - 2n(a + 1)}{(2n + 1)(a + 1)}, ~ ~ ~ s = \frac{N}{(2n + 1)(a + 1)}.
$$
Integrating Eq. (\ref{eq38}), we obtain
\begin{equation}
\label{eq39}
g = \beta \sinh^{\frac{1}{(1 + r)}}(b t + t_{0}),
\end{equation}
where $\beta = (c_{1})^{\frac{1}{1 + r}}$, $b = \sqrt{s(1 + r)}$ and $t_{0}$, $c_{1}$ are constants 
of integration. \\
Thus from (\ref{eq36}) we get
\begin{equation}
\label{eq40}
k = \delta \beta^{a}\sinh^{\frac{a}{(1 + r)}}(b t + t_{0}). 
\end{equation}
Eqs. (\ref{eq33}) and (\ref{eq37}) lead to
\begin{equation}
\label{eq41}
hh_{11} + \ell h^{2}_{1} = m h^{2},
\end{equation}
where
$$ \ell = \frac{K(K - 1) - 2n(K + 1)}{(2n + 1)(K + 1)}, ~ ~ m = \frac{N}{(2n + 1)(K + 1)}.$$
Integrating Eq. (\ref{eq41}), we obtain
\begin{equation}
\label{eq42}
h = r_{0}\sinh^{\frac{1}{(1 + \ell)}}(cx + x_{0}),
\end{equation}
where $r_{0} = c_{2}^{\frac{1}{1 + \ell}}$, $c = \sqrt{m(1 + \ell)}$ and $c$, $x_{0}$ are constants of 
integration. \\
Hence from (\ref{eq35}) and (\ref{eq42}) we get
\begin{equation}
\label{eq43}
f = \alpha r^{K}_{0} \sinh^{\frac{K}{(1 + \ell)}}(cx + x_{0}).
\end{equation}
Hence, we obtain
\begin{equation}
\label{eq44}
B = fg = Q\sinh^{\frac{K}{(\ell + 1)}}(cx + x_{0})\sinh^{\frac{1}{(r + 1)}}(bt + t_{0}),
\end{equation}
\begin{equation}
\label{eq45}
C = hk = R\sinh^{\frac{1}{(\ell + 1)}}(cx + x_{0})\sinh^{\frac{a}{(r + 1)}}(bt + t_{0}),
\end{equation}
\begin{equation}
\label{eq46}
A = (BC)^{n} = M\sinh^{\frac{n(K + 1)}{(\ell + 1)}}(cx + x_{0})\sinh^{\frac{n(a + 1)}{(r + 1)}}
(bt + t_{0}),
\end{equation}
where
$ Q = \alpha \beta r^{K}_{0}$, $R = r_{0} \delta \beta^{a}$, $M = (QR)^{n}$. \\

After using suitable transformation of coordinates metric (\ref{eq1}) reduces to
\[
ds^{2}= M^{2}\sinh^{\frac{2n(K + 1)}{(\ell + 1)}}(c X) sinh^{\frac{2n(a + 1)}{(r + 1)}} (b T) 
(dX^{2} - dT^{2}) + 
\]
\begin{equation}
\label{eq47}
Q^{2} \sinh^{\frac{2K}{(\ell + 1)}}(c X)\sinh^{\frac{2}{(r + 1)}}(b T)dY^{2} + R^{2} 
\sinh^{\frac{2}{(\ell + 1)}}(c X)\sinh^{\frac{2a}{r + 1}}(b T)dZ^{2}, 
\end{equation}
where $X = x + \frac{x_{0}}{c}$, $Y = Qy$, $Z = Rz$ and $T = t + \frac{t_{0}}{b}$. \\

The energy density $(\rho)$, the string tension density $(\lambda)$, the particle 
density $(\rho_{p})$, the scalar of expansion $(\theta)$, shear tensor $(\sigma)$, 
acceleration vector $\dot{u}_{i}$ and the 
proper volume for $(V^{3})$ for the model (\ref{eq47}) are given by 
\[
\rho = \frac{1}{M^{2}\sinh^{\frac{2n(K + 1)}{(\ell + 1)}}(cX) \sinh^{\frac{2n(a + 1)}{(r + 1)}}
(b T)}\times
\]
\[
\Biggl[\frac{c^{2}\{(K + 1)(n(K + 1) + \ell) - K^{2}\}}{(\ell + 1)^{2}}\coth^{2}(c X) 
\]
\[
+ \frac{b^{2}\{n(a + 1)^{2} + a\}}{(r + 1)^{2}}\coth^{2}(bT) - \frac{c^{2}(K + 1)}{(\ell + 1)}
\]
\begin{equation}
\label{eq48}
+ \frac{F^{2}_{12}}{Q^{2}\bar{\mu}\sinh^{\frac{2K}{(\ell + 1)}}(cX) \sinh^{\frac{2}{(r + 1)}}(bT)}\Biggr],
\end{equation}
\[
\lambda = \frac{1}{M^{2}\sinh^{\frac{2n(K + 1)}{(\ell + 1)}}(c X) \sinh^{\frac{2n(a + 1)}{(r + 1)}}
(b T)}\times
\]
\[
\Biggl[\frac{b^{2}\{a^{2} - (a + 1)(na + n + r)\}}{(r + 1)^{2}}\coth^{2}(bT) 
\]
\[
- \frac{c^{2}\{n(K + 1)^{2} - K\}}{(\ell + 1)^{2}}\coth^{2}(cX) + \frac{b^{2}(a + 1)}{(r + 1)}
\]
\begin{equation}
\label{eq49}
+ \frac{F^{2}_{12}}{Q^{2}\bar{\mu}\sinh^{\frac{2K}{(\ell + 1)}}(cX) 
\sinh^{\frac{2}{(r + 1)}}(bT)}\Biggr],
\end{equation}
\[
\rho_{p} = \rho - \lambda = \frac{1}{M^{2}\sinh^{\frac{2n(K + 1)}{(\ell + 1)}}(cX) 
\sinh^{\frac{2n(a + 1)}{(r + 1)}}(b T)}\times
\]
\[
\Biggl[\frac{b^{2}\{(a + 1)(2na + 2n + r) + a(1 -a)\}}{(r + 1)^{2}}\coth^{2}(bT)
\]
\begin{equation}
\label{eq50}
+ \frac{c^{2}(K + 1)\{(2nK + 2n + \ell) - K\}}{(\ell + 1)^{2}}\coth^{2}(cX) - 
\frac{c^{2}(K + 1)}{\ell + 1)} - \frac{b^{2}(a + 1)}{(r + 1)}\Biggr]
\end{equation}
where
\[
F^{2}_{12} = \bar{\mu} Q^{2}\sinh^{\frac{2K}{(\ell + 1)}}(cX) \sinh^{\frac{2}{(r + 1)}}(bT)
\Biggl[\frac{b^{2}(1 - a)}{(r + 1)} + \frac{c^{2}(1 - K)}{(\ell + 1)} - 
\]
\begin{equation}
\label{eq51}
\frac{b^{2}[r + a(a - r - 1)]}{(r + 1)^{2}} \coth^{2}(bT) - \frac{c^{2}[\ell + K(K - \ell - 1)]}
{(\ell + 1)^{2}}\coth^{2}(c X)\Biggr], 
\end{equation}
\begin{equation}
\label{eq52}
\theta = \frac{b(a + 1)(n + 1) \coth(bT)}{(r + 1)M\sinh^{\frac{n(K + 1)}{(\ell + 1)}}(c X) 
\sinh^{\frac{n(a + 1)}{(r + 1)}}(b T)},
\end{equation}
\begin{equation}
\label{eq53}
\sigma^{2} = \frac{b^{2}\{(a + 1)^{2}(n^{2} - n + 1) - 3a\} \coth^{2}(bT)}{3(r + 1)^{2}M^{2} 
\sinh^{\frac{2n(K + 1)}{(\ell + 1)}}(c X) \sinh^{\frac{2n(a + 1)}{(r + 1)}}(b T)},
\end{equation}
\begin{equation}
\label{eq54}
\dot{u_{i}} = \left(\frac{c n(K + 1)}{(\ell + 1)}\coth (cX), 0, 0, 0\right),
\end{equation}
\begin{equation}
\label{eq55}
V^{3} = (QR)^{(2n + 1)} \sinh^{\frac{(2n + 1)(K + 1)}{(\ell + 1)}}(cX) \sinh^{\frac{(2n + 1)(a + 1)}{(r + 1)}}(bT).
\end{equation}
From Eqs. (\ref{eq52}) and (\ref{eq53}) we obtain
\begin{equation}
\label{eq56}
\frac{\sigma^{2}}{\theta^{2}} = \frac{(a + 1)^{2}(n^{2} - n + 1) - 3a}{3(n + 1)^{2}
(a + 1)^{2}} = \mbox{(constant)}.
\end{equation}
The deceleration parameter $(q)$ in presence of magnetic field is given by 
\[
q = -1 + \frac{3(r + 1)M\sinh^{\frac{n(K + 1)}{(\ell + 1)}}(cX) \sinh^{\frac{n(a + 1)}{(r + 1)}}(bT)}
{b(a + 1)(n + 1)\coth(bT)}\times
\]
\begin{equation}
\label{eq57}
\left[\frac{2b}{\sinh(bT)} + \frac{nb(a + 1)}{(r + 1)}\coth(bT)\right].
\end{equation}
\section{Conclusions}
In this paper, we have investigated the behaviour of a string in the cylindrically 
symmetric inhomogeneous cosmological model with electromagnetic field.

If we choose the suitable values of constants $K$ and $M$, we find that energy 
conditions $\rho \geq 0$, $\rho_{p} \geq 0$ are satisfied. The string tension 
$(\lambda)$ and energy density $(\rho)$ increases as $F_{12}$ increases. 

The model (\ref{eq47}) starts with a big bang at $T = 0$. The expansion in the 
model decreases as time increases . The expansion in the model stops at $T = \infty$. 
Since $ \frac{\sigma}{\theta} \ne 0$, hence the model does not 
approach isotropy in general. However, $ (a + 1)^{2}(n^{2} - n + 1) - 3a = 0$, then 
$ \frac{\sigma}{\theta} = 0$ which leads the isotropy of the universe. We also observe that 
$\rho$, $\lambda$, $\rho_{p}$ tend to $\infty$ when $X \to 0$, $T \to 0$.  
The energy density $(\rho)$ and string tension density $(\lambda)$ increases as electromagnetic 
field component $(F_{12})$ increases. The proper volume $V^{3}$ increases exponentially as 
time increases. Thus, the model represents the inflationary scenario. The model  (\ref{eq47}) 
has a point type singularity at $T = 0$ (MacCallum \cite{ref44}). 

We observe that $q < 0$ if
\[
\frac{3(r + 1)M\sinh^{\frac{n(K + 1)}{(\ell + 1)}}(cX) \sinh^{\frac{n(a + 1)}{(r + 1)}}(bT)}
{b(a + 1)(n + 1)\coth(bT)}\times
\]
\[
\left[\frac{2b}{\sinh(bT)} + \frac{nb(a + 1)}{(r + 1)}\coth(bT)\right] < 0.
\]
The deceleration parameter $q$ approaches the value $(-1)$ as in the case of de-Sitter universe if
\[
2(r + 1) + n(a + 1)\cosh(bT) = 0.
\]


\begin{thebibliography}{000}
\bibitem {ref1}
Ya. B. Zel'dovich, I. Yu. Kobzarev and L. B. Okun, Zh. Eksp. Teor. Fiz. {\bf 67}, 3 (1975); 
Sov. Phys.-JETP {\bf 40}, 1 (1975).  
\bibitem {ref2}
T. W. B. Kibble, J. Phys. A: Math. Gen. {\bf 9}, 1387 (1976).
\bibitem {ref3}  
T. W. B. Kibble, Phys. Rep. {\bf 67}, 183 (1980).
\bibitem {ref4}
A. E. Everett, Phys. Rev. {\bf 24}, 858 (1981).
\bibitem {ref5}
A. Vilenkin, Phys. Rev. D {\bf 24}, 2082 (1981).
\bibitem {ref6}
Ya. B. Zel'dovich, Mon. Not. R. Astron. Soc. {\bf 192}, 663 (1980).
\bibitem {ref7}
P. S. Letelier, Phys. Rev. D {\bf 20}, 1249 (1979). 
\bibitem {ref8}
P. S. Letelier, Phys. Rev. D {\bf 28}, 2414 (1983).
\bibitem {ref9}
J. Stachel, Phys. Rev. D {\bf 21}, 2171 (1980).
\bibitem {ref10} 
Ya. B. Zel'dovich, A. A. Ruzmainkin, and D. D. Sokoloff,  
{\it Magnetic field in Astrophysics}, Gordon and Breach, New Yark, (1983). 
\bibitem {ref11} 
E. R. Harrison, Phys. Rev. Lett. {\bf 30}, 188 (1973). 
\bibitem {ref12}  
H. P. Robertson and A. G. Walker, Proc. London Math. Soc. {\bf 42}, 90 (1936). 
\bibitem {ref13} 
C. W. Misner, K. S. Thorne, and J. A. Wheeler, {\it Gravitation}, W. H.
Freeman, New York, (1973).  
\bibitem {ref14}  
E. Asseo and H. Sol, Phys. Rep. {\bf 6}, 148 (1987).
\bibitem {ref15}  
R. Pudritz and J. Silk, Astrophys. J. {\bf 342}, 650 (1989).
\bibitem {ref16}  
K. T. Kim, P. G. Tribble, and P. P. Kronberg, Astrophys. J. {\bf 379}, 80 (1991).
\bibitem {ref17}  
R. Perley and G. Taylor, Astrophys. J. {\bf 101}, 1623 (1991).
\bibitem {ref18}  
P. P. Kronberg, J. J. Perry, and E. L. Zukowski, Astrophys. J. {\bf 387}, 528 (1991).
\bibitem {ref19}  
A. M. Wolfe, K. Lanzetta, and A. L. Oren, Astrophys. J. {\bf 388}, 17 (1992).
\bibitem {ref20}  
R. Kulsrud, R. Cen, J. P. Ostriker, and D. Ryu, Astrophys. J. {\bf 380}, 481 (1997).   
 \bibitem {ref21}  
J. D. Barrow, Phys. Rev. D {\bf 55}, 7451 (1997).
\bibitem {ref22} 
M. A. Melvin, Ann. New York Acad. Sci. {\bf 262}, 253 (1975).
\bibitem {ref23}
A. Banerjee, A. K. Sanyal, and S. Chakraborty, Pramana-J. Phys. {\bf 34}, 1 (1990).
\bibitem {ref24}
S. Chakraborty, Ind. J. Pure Appl. Phys. {\bf 29}, 31 (1980).
\bibitem {ref25}
R. Tikekar and L. K. Patel, Gen. Rel. Grav. {\bf 24}, 397 (1992).
\bibitem {ref26} 
R. Tikekar and L. K. Patel, Pramana-J. Phys. {\bf 42}, 483 (1994). 
\bibitem {ref27}
L. K. Patel and S. D. Maharaj, Pramana-J. Phys. {\bf 47}, 1 (1996).
\bibitem {ref28}
S. Ram and T. K. Singh, Gen. Rel. Grav. {\bf 27}, 1207 (1995).
\bibitem {ref29}
J. Carminati and C. B. G. McIntosh, J. Phys. A: Math. Gen. {\bf 13}, 953 (1980).
\bibitem {ref30}
G. P. Singh and T. Singh, Gen. Rel. Grav. {\bf 31}, 371 (1999).
\bibitem {ref31}
G. P. Singh, Nuovo Cim. B {\bf 110}, 1463 (1995); Pramana-J. Phys. {\bf 45}, 189 (1995).
\bibitem {ref32}
J. E. Lidsey, D. Wands, and E. J. Copeland, Phys. Rep. {\bf 337}, 343 (2000).
\bibitem {ref33}
R. Bali and R. D. Upadhaya, Astrophys. Space Sci. {\bf 283}, 97 (2003).
\bibitem {ref34}
R. Bali and Anjali, Astrophys. Space Sci. {\bf 302}, 201 (2006). 
\bibitem {ref35}
R. Bali, U. K. Pareek and A. Pradhan, Chin. Phys. Lett. {\bf 24}, (2007), To appear.
\bibitem {ref36}
R. Bali and A. Tyagi, Gen. Rel. Grav. {\bf 21}, 797 (1989).
\bibitem {ref37}  
A. Pradhan, I. Chakrabarty, and N. N. Saste, Int. J. Mod. Phys. D {\bf 10}, 741 (2001).
\bibitem {ref38}  
A. Pradhan, P. K. Singh, and K. Jotania, Czech. J. Phys. {\bf 56}, 641 (2006).
\bibitem {ref39}
J. D. Barrow and K. E. Kunze, Phys. Rev. D {\bf 56}, 741 (1997).
\bibitem {ref40}
J. D. Barrow and K. E. Kunze, Phys. Rev. D {\bf 57}, 2255 (1998).
\bibitem {ref41}
H. Baysal, I. Yavuz, I. Tarhan, U. Camci, and I. Yilmaz, Turk. J. Phys. {\bf 25}, 
283 (2001).
\bibitem {ref42}
A. Lichnerowica, {\it Relativistic Hydrodynamics and Magnetohydrodynamics}, W. A. Benjamin 
Inc., New York, p. 93 (1967).
\bibitem {ref43}
J. L. Synge, {\it Relativity: The General Theory}, North-Holland Publ. Co., 
Amsterdam, p. 356 (1960).
\bibitem {ref44}
M. A. H. MacCallum, Comm. Math. Phys. {\bf 20}, 57 (1971).
\end{thebibliography}
\end{document}